\documentclass{llncs}
\usepackage[table,xcdraw,dvipsnames]{xcolor}
\usepackage[T1]{fontenc}
\usepackage{utfsym}
\usepackage{xcolor} % 加载xcolor宏包以支持颜色
\usepackage{fontawesome} % 加载fontawesome宏包以支持图标
\usepackage[normalem]{ulem}
\useunder{\underline}{\ul}{}
\usepackage[table,xcdraw,dvipsnames]{xcolor} 
\usepackage{booktabs} % 提供更好的表格线条
\usepackage{graphicx} % 提供\resizebox命令
\usepackage[table,xcdraw,dvipsnames]{xcolor} % 提供表格颜色设置
\usepackage[normalem]{ulem} 
\usepackage{booktabs}
\usepackage{multirow}
\usepackage{amsmath}
\usepackage{amssymb}
\usepackage{booktabs} % 用于表格的美化
\usepackage{array}    % 用于自定义列格式
\usepackage{multirow}
\usepackage{makecell} % 用于单元格内容的对齐
\usepackage{graphicx} % 用于调整表格大小
\usepackage{bbding}
\usepackage{tabularx}
\definecolor{forestgreen}{rgb}{0.13, 0.55, 0.13} 
\usepackage{graphicx,verbatim}
\usepackage{caption} % 在导言区添加此包
\usepackage{graphicx} % 用于缩放表格
\usepackage{booktabs} % 用于美化表格线条
\usepackage{caption} % 用于添加表格标题
\usepackage{tabularx} % 用于自适应宽度的表格
\usepackage{multirow} % 用于合并单元格
\usepackage[colorlinks,urlcolor=blue,citecolor=blue]{hyperref}

\usepackage{marvosym}%

\begin{document}
%
% \title{From Vision to Text CoT: Advancing Medical Diagnosis with Multimodal Language Models}

% \title{From Vision to Text CoT: Advancing Medical Diagnosis with Multimodal Language Models}

% \title{From Vision to Text CoT: Enhancing Medical Diagnosis through Multimodal Language Model}

\title{V2T-CoT: From Vision to Text Chain-of-Thought for Medical Reasoning and Diagnosis}
\titlerunning{V2T-CoT}

% \title{V2T-CoT: Enhancing Medical Diagnosis Reasoning with Joint Vision and Text Chain-of-Thought}
% \title{V2T-CoT: Integrating Region-Level Localization and Multimodal Language Models for Medical Diagnosis}
%
% \begin{comment}  %% Removed for anonymized MICCAI 2025 submission
\author{Yuan Wang\inst{1}\thanks{Equal contribution. \textsuperscript{\Letter}Corresponding author.} %1{Wang, Yuan}
\and Jiaxiang Liu\inst{1}$^{*}$  %2{Liu, Jiaxiang}
\and Shujian Gao\inst{2} %3{Gao, Shujian}
\and Bin Feng \inst{3} %4{Feng, Bin}
\and Zhihang Tang \inst{4} %5{Tang, Zhihang}
\and Xiaotang Gai \inst{1} %6{Gai, Xiaotang}
\and Jian Wu \inst{1} \and %7{Wu, Jian}
Zuozhu Liu \inst{1}$^{\href{mailto:zuozhuliu@intl.zju.edu.cn}{\textrm{\Letter}}}$} %8{Liu, Zuozhu}

\authorrunning{Yuan Wang, Jiaxiang Liu, et al.}
% First names are abbreviated in the running head.
% If there are more than two authors, 'et al.' is used.
%
\institute{Zhejiang Key Laboratory of Medical Imaging Artificial Intelligence, Zhejiang University, HangZhou, China \and
Academy for Engineering and Technology, Fudan University, Shanghai, China \and
Department of Oral and Maxillofacial Radiology, Stomatology Hospital, School of Stomatology,Zhejiang University, Hangzhou, China \and 
Intelligent Computing Infrastructure Innovation Center, Zhejiang Lab, Hangzhou, China\\
\email{\{yuan2.24,zuozhuliu\}@intl.zju.edu.cn}\\
\url{https://github.com/Venn2336/V2T_CoT}}

% \end{comment}

% Table 1 Setting
% \renewcommand{\arraystretch}{1.1}  % 增加行间距
% \setlength{\tabcolsep}{10pt}  % 增加列间距

% \author{Anonymized Authors}  %% Added for anonymized MICCAI 2025 submission
% \authorrunning{Anonymized Author et al.}
% \institute{Anonymized Affiliations \\
%     \email{email@anonymized.com}}

\maketitle              

%Recent advancements in Medical Visual Question Answering (Med-VQA) have led to significant progress in the field. However, most existing models focus on global image features rather than localizing disease-specific regions crucial for diagnosis. While current methods prioritize answer accuracy, they neglect the reasoning pathways essential for clinical trustworthiness. We propose V2T-CoT: Vision-to-Text Chain-of-Thought, which introduces two novel components: 1) an automated lesion localization module generating Vision CoT through regional pixel attention, and 2) Text CoT providing stepwise diagnostic rationale. To enable multimodal reasoning, we construct R-Med 39K - the first medical instruction dataset containing 39K vision-language pairs with expert-validated reasoning chains. By jointly optimizing visual grounding and textual rationale generation, our framework achieves 84.86\% accuracy on VQA-RAD and 87.61\% on SLAKE, outperforming state-of-the-art methods by 2.39-4.78\% while producing interpretable lesion maps. Code and datasets will be publicly released.

\begin{abstract}
Recent advances in multimodal techniques have led to significant progress in Medical Visual Question Answering (Med-VQA).
However, most existing models focus on global image features rather than localizing disease-specific regions crucial for diagnosis.
Additionally, current research tends to emphasize answer accuracy at the expense of the reasoning pathway, yet both are crucial for clinical decision-making.
To address these challenges, we propose From Vision to Text Chain-of-Thought (\textbf{V2T-CoT}), a novel approach that automates the localization of preference areas within biomedical images and incorporates this localization into region-level pixel attention as knowledge for Vision CoT. 
By fine-tuning the vision language model on constructed \textbf{R-Med 39K} dataset, V2T-CoT provides definitive medical reasoning paths. V2T-CoT integrates visual grounding with textual rationale generation to establish precise and explainable diagnostic results. 
Experimental results across four Med-VQA benchmarks demonstrate state-of-the-art performance, achieving substantial improvements in both performance and interpretability.

% \footnote{Samples in anonymous repository: https://anonymous.4open.science/r/R-Med39K}


\end{abstract}
%
% \footnote{\textsuperscript{$^{*}$}Equal contribution. \textsuperscript{\Letter}Corresponding author.}

%
%
\section{Introduction}

Medical Visual Question Answering (Med-VQA) has emerged as a critical domain in healthcare AI, leveraging multimodal deep learning to answer clinical questions related to images posed in natural language \cite{chen2022align,gong2021cross,he2024pefomed,jiang2025omnivmedscalingmedicalvisionlanguage}. Med-VQA holds promise for automated, personalized health consultations, addressing the growing demand for accessible and timely healthcare \cite{liu2023parameter}. By integrating textual data (e.g., clinical notes, patient histories) with medical images (e.g., X-rays, MRIs), these models provide a holistic understanding of patient conditions, improving decision-making and care delivery.

\begin{figure}[t]
    \centering
    \includegraphics[width = \textwidth]{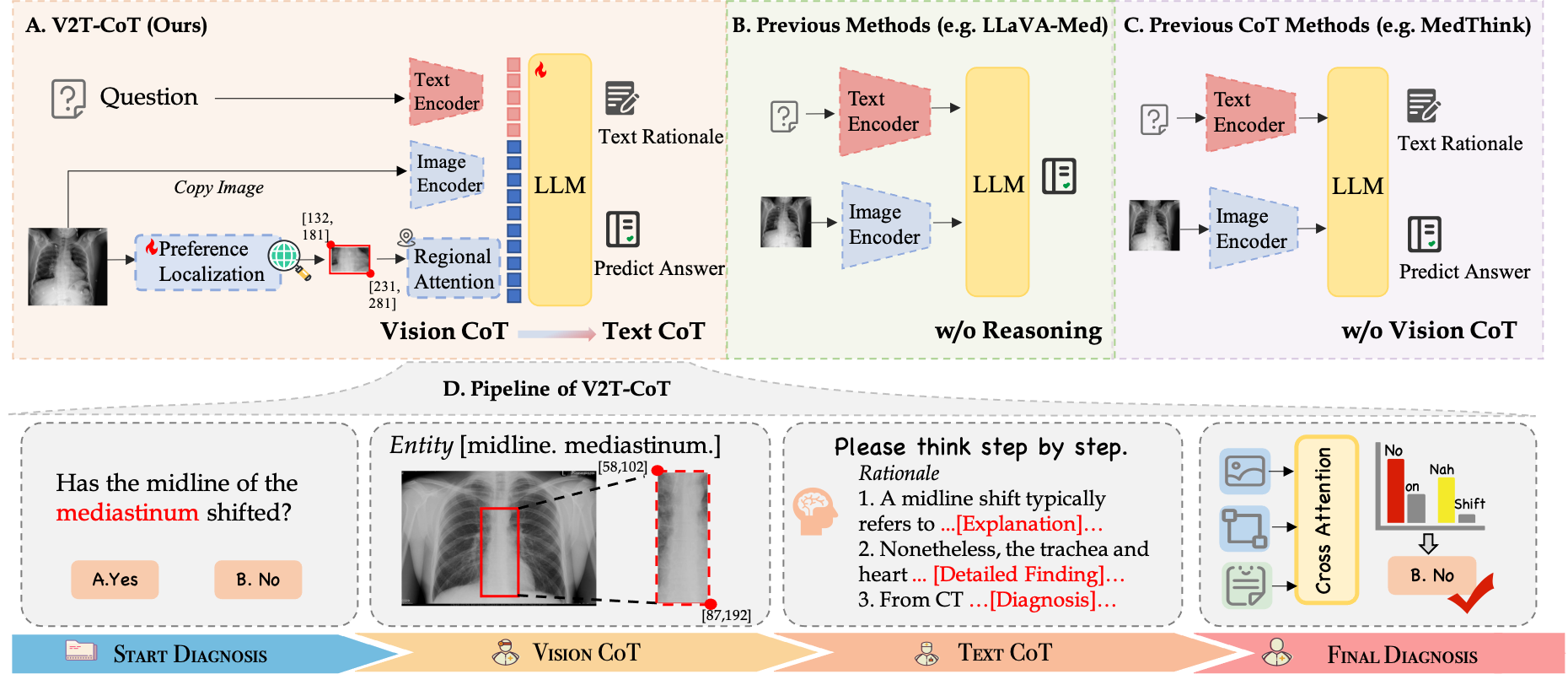}
    \caption{The comparison between V2T-CoT and existing Med-VQA methods. 
    \textbf{A} employs a combined vision and text encoding strategy with regional attention for medical diagnosis. In contrast, previous methods (\textbf{B} \& \textbf{C}) either lack reasoning or utilize it in a text-only context. \textbf{D} demonstrates the pipeline of V2T-CoT.}
    \label{pipeline1}
\end{figure}

Recent advancements in Med-VQA have adopted deep learning frameworks such as DoctorGLM \cite{xiong2023doctorglm},Huatuo-o1 \cite{chen2024huatuogpto1medicalcomplexreasoning} and MedFound \cite{liu2025generalist} to enhance the medical reasoning process. However, three core challenges persist \cite{zhang2023multimodal,liu-etal-2024-medcot,zheng2023ddcot,jiang2025capo}: \textbf{First}, existing feature fusion mechanisms (e.g., concatenation/weighted operations) suffer from representational capacity limitations. These coarse-grained fusion strategies fail to establish precise mappings between visual key regions (e.g., lesion areas in CT/MRI) and textual semantics, resulting in ineffective capture of diagnosis-relevant fine-grained cross-modal correlations. \textbf{Second}, high-accuracy models lack explainable reasoning pathways. Taking LLaVA-Med \cite{li2024llava} (Fig.~\ref{pipeline1}\textcolor{red}{B}) as an example, despite achieving high accuracy, its black-box decision mechanism cannot generate clinical reasoning chains (e.g., progressive analysis for malignancy determination based on tumor morphological features), severely hindering trustworthiness verification in diagnostic scenarios \cite{liu-etal-2024-medcot}. \textbf{Third}, multimodal CoT methods face both data and algorithm constraints. Frameworks like MedThink \cite{gai2024medthink} (Fig.~\ref{pipeline1}\textcolor{red}{C}), while incorporating clinical reasoning chains, are limited by: annotation data scarcity due to manual labeling costs; absence of specialized instruction-tuning datasets for training reasoning paths \cite{savage2024diagnostic,liu-etal-2024-medcot}. These limitations collectively lead to insufficient generalization capabilities in multi-perspective complex medical analysis.

In this paper, we propose a novel Med-VQA method: From Vision to Text CoT (\textbf{V2T-CoT}), which addresses the aforementioned challenges by effectively integrating textual and visual reasoning (Fig.~\ref{pipeline1}\textcolor{red}{A}). This method combines visual clues with textual information, comprising two core components: (1) a preference localization mechanism that identifies key regions in descriptive text relevant to the image; (2) a regional pixel-level attention mechanism that focuses on specific areas of medical images crucial for final diagnosis. To further enable explainable reasoning, we construct an instruction-tuning dataset \textbf{R-Med 39K} containing multi-granularity reasoning paths. Ultimately, our vision language model (VLM) generates coherent and interpretable diagnostic rationales, from which medical conclusions are derived. Experiments demonstrate that this approach achieves SOTA performance under comparable parameter sizes, significantly outperforming existing models. The main contributions are summarized as follows:

\begin{itemize}
    \item We propose a medical visual to text chain-of-thought reasoning framework, \textbf{V2T-CoT}, which provides disease-related visual cues and a clear reasoning path, facilitating medical diagnosis.
    \item To locate disease-related visual cues, V2T-CoT introduces an automated method (Vision CoT) for identifying and aligning relevant image regions.
    \item We constructed \textbf{R-Med 39K} (Instruction-Tuning dataset) with reasoning paths from four Med-VQA datasets for Text CoT, integrating them into VLM training for reliable rationale validated by both LLMs and experts.
    \item Through extensive experiments, we validated the superior performance of V2T-CoT on four Med-VQA datasets compared to existing methods, demonstrating the effectiveness of Vision CoT for visual localization and the rationale in Text CoT for reasoning pathways, meeting the clarity and interpretability requirements of both humans and LLMs.
\end{itemize}

\section{Method}
\subsection{Overview}

In this work, we propose a novel multimodal approach for Med-VQA that integrates visual and textual reasoning chains. As shown in Fig.~\ref{pipeline1}\textcolor{red}{D}, V2T-CoT first leverages a vision encoder to extract spatial-semantic features from critical anatomical regions, explicitly capturing structural deviations such as midline shifts (Vision CoT). The reasoning module then synthesizes anatomical region proposals with multimodal evidence to construct diagnostic rationales (Text CoT), ultimately delivering clinically interpretable diagnoses with both accuracy and clinical diagnostic transparency.

\subsection{Vision CoT}

\subsubsection{Formulation of Phrase Grounding.} 

\begin{figure}[t]
    \centering
    \includegraphics[width = \textwidth]{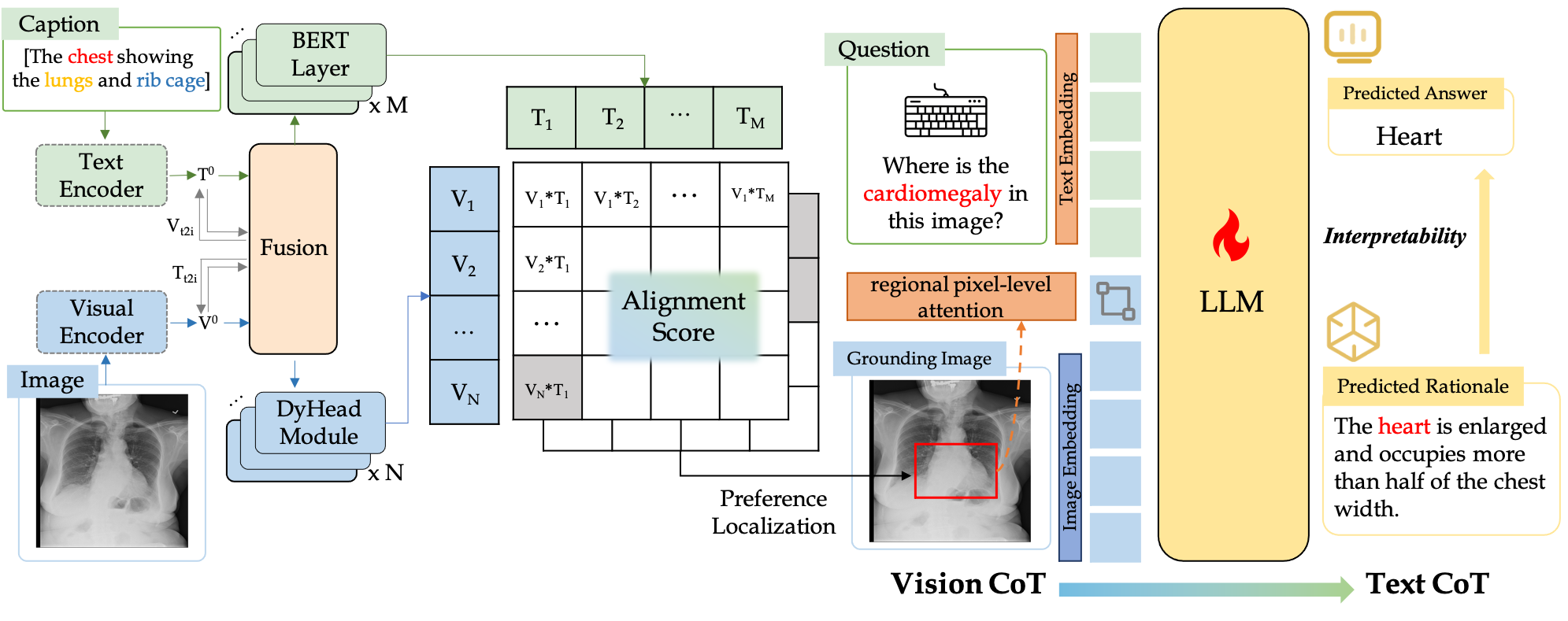}
    \caption{Implementation of V2T-CoT for medical diagnosis. The model leverages dynamic fusion of image and text encodings, alignment scoring, and regional attention mechanisms to generate answers and rationale.}
    \label{pipeline3}
\end{figure}

We reformulate medical object detection as phrase grounding within the Vision CoT framework, aligning each image region with text prompt phrases. As shown in Fig.~\ref{pipeline3}, given an image and text prompt, visual encoder $\text{Enc}_{I}$ extracts region features $V \in \mathbb{R}^{N\times d}$, while language encoder $\text{Enc}_{L}$ encodes textual tokens $T \in \mathbb{R}^{M \times d}$. The alignment score matrix $S_{\text{ground}} \in \mathbb{R}^{N\times M}$ is computed via:
$S_{\text{ground}} = V T^{\top}$,
where $V$ and $T$ denote region-level visual features and phrase-level text embeddings respectively.

\subsubsection{Fusion of Visual and Language Features.} 
To enhance the interaction between visual and textual modalities, we introduce a fusion mechanism in the last layers of the encoders.
This mechanism involves cross-modality multi-head attention (X-MHA) modules that facilitate the exchange of information between the image and text encoders. The fused encoder is defined as:
\begin{align}
    & V^i_{\text{t2i}}, T^i_{\text{i2t}} = \text{X-MHA}(V^i, T^i), \quad i \in \{0, 1, .., L-1\}, \label{eq:xmha} \\
    & V^{i+1} = \text{DyHeadModule}(V^i + V^i_{\text{t2i}}), \quad V = V^{L}, \label{eq:i2t} \\
    & T^{i+1} = \text{BERTLayer}(T^i + T^i_{\text{i2t}}), \quad T = T^{L}, \label{eq:t2i}
\end{align}
where $L$ is the number of DyHeadModules, $V^i_{\text{t2i}}$ and $T^i_{\text{i2t}}$ are the context vectors generated by the X-MHA module.

\subsubsection{Regional Pixel-Level Attention.}
Following Preference Localization, we refine the model's focus through regional pixel-level attention, enhancing its ability to target specific areas crucial for diagnosing conditions such as cardiomegaly.

Regional pixel-level attention is implemented by assigning higher weights to key pixels using convolutional layers and attention modules. To balance regional and global attention, we introduce a weight factor \(\alpha\), where \(\alpha \in [0, 1]\). The combined attention \(A_{\text{combined}}\) is computed as:

\begin{align}
A_{\text{combined}} = \text{SoftMax} \left( \alpha \cdot Q_{\text{regional}} K_{\text{regional}}^{\top} + (1 - \alpha) \cdot Q_{\text{global}} K_{\text{global}}^{\top} \right) / \sqrt{d},
\end{align}
where \(Q_{\text{regional}} \in \mathbb{R}^{h \times w \times d}\) and \(K_{\text{regional}} \in \mathbb{R}^{h \times w \times d}\) are the query and key features from the localized region, while \(Q_{\text{global}} \in \mathbb{R}^{H \times W \times d}\) and \(K_{\text{global}} \in \mathbb{R}^{H \times W \times d}\) are the query and key features from the global image. The weight \(\alpha\) controls the balance between the local and global attention, allowing the model to focus on fine-grained details while considering broader contextual information.

\subsection{Text CoT}

\subsubsection{R-Med 39K Dataset Construction.} 
\label{Text Rationale}
The text rationale construction process adopts a two-stage generation and verification pipeline. Initially, given the input tuple $<Question(Q),Image(I)>$, a vision language model generates the preliminary answer(A). Subsequently, leveraging the triplet $<Q,I,A>$, we employ a reasoning prompting strategy to produce a structured diagnostic rationale(R). To ensure clinical validity, the generated R undergoes verification where a separate language model evaluates its logical consistency with A and I. This process yields the final quadruple $<Q,I,A,R>$ for instruction tuning. As demonstrated in Figure \ref{fig:judgement}, quality control combines automated metrics with expert validation, with implementation details provided in Section \ref{Text CoT Rationale}.

\subsubsection{Implementation of the Text CoT.} 

Given the visual features \( V \in \mathbb{R}^{N \times d} \) extracted from the Vision CoT and the textual features \( T \in \mathbb{R}^{M \times d} \) encoded by the language model, the Text CoT module computes the cross-modal alignment scores as: $
\small
S_{\text{text}} = \text{SoftMax} \left( \frac{Q_{\text{text}} K_{\text{text}}^{\top}}{\sqrt{d}} \right),
$
where \( Q_{\text{text}}\) and \( K_{\text{text}}\) are the query and key projections of the textual and visual features, respectively.

To further enhance the interaction between modalities, we employ a multi-head cross-attention mechanism in the VLM, where the textual features are iteratively refined through layers of self-attention:
\begin{align}
T^{i+1} = \text{MultiHeadAttention}(T^{i}, V_{regional}, V_{global}),
\end{align}
where \( T^{i} \) represents the textual features at layer \( i \). This iterative refinement ensures that the textual reasoning aligns with the visual clues.

\section{Experiments}
\subsection{Experiments Setting}
\subsubsection{Datasets.} 
Four benchmarks are used as the training and testing datasets for V2T-CoT. 
Among them, VQA-RAD \cite{lau2018dataset}, SLAKE \cite{liu2021slake}, and VQA-2019 \cite{ben2019vqa} are based on radiology images (e.g., CT, MRI), while PathVQA \cite{he2020pathvqa} is built on pathological images (e.g., tissue slides). Following Section \ref{Text Rationale}, we generated rationales (thinking pathway) via LLMs (GPT-4 and Gemini), and construct them into instruction-tuning dataset R-Med containing 39K quadruple$<Q,I,A,R>$ format.
In accordance with the original data partition, the questions are categorized into closed-ended (single-answer) and open-ended (free-form) types \cite{liu-etal-2024-medcot}.

\subsubsection{Implementation Details.} 

Vision CoT initializes with the GLIP \cite{9879567} detector, fine-tuned on SLAKE to adapt it to the medical domain. It is then deployed in an automated pipeline to generate region proposals for other datasets without manual annotation.
For VLM, we employ CLIP-ViT-L/14@336px as the vision encoder to extract features from input images, distinguishing between global attention and regional pixel-level attention.
For the language model, StableLM (1.6B) and Phi2 (2.7B) are utilized \cite{bellagente2024stablelm216b,gunasekar2023textbooksneed}. 
Additionally, a 2-layer Multi-Layer Perceptron (MLP) is implemented as the projector to facilitate the integration of visual and textual features.

In the pretraining phase, the 2-layer multimodal projector is trained on a large-scale dataset of medical image-caption pairs \cite{li2024llava}, establishing a foundational understanding of visual data \cite{liu2024visual}. 

In the instruction fine-tuning phase, the CLIP encoder is frozen, and the remaining model components, including the visual encoder and language model, are fine-tuned on our R-Med dataset consisting of up to 39K quadruple. 
Both pretraining and fine-tuning are performed on 4 NVIDIA GeForce RTX 3090 GPUs.

\begin{table*}[htbp!]
\large
\setlength{\abovecaptionskip}{10pt}  % 在特定表格中调整
\centering
\caption{Comparison of model performances across different datasets on closed-end question: The \textbf{bolded} and \underline{underlined} values correspond to the best and 2nd best performance, respectively. All results are in \%.}
\resizebox{\textwidth}{!}{ % 调整表格宽度为页面宽度
\begin{tabular}{lcccccc}
\rowcolor[HTML]{EFEFEF} 
\textbf{Method}                    & \multicolumn{1}{l|}{\textbf{w/ MLLM}}                                              & \multicolumn{1}{c|}{\textbf{Act.}} & \multicolumn{1}{c|}{\textbf{VQA-RAD}} & \multicolumn{1}{c|}{\textbf{SLAKE}} & \multicolumn{1}{c|}{\textbf{VQA-2019}} & \textbf{PathVQA} \\
\midrule
\multicolumn{7}{l}{Representative \& SoTA methods with numbers reported in the literature} \\
\midrule
VL Encoder-Decoder \cite{bazi2023vision}          & \multicolumn{1}{c|}{}                                                               & \multicolumn{1}{c|}{-}            & \multicolumn{1}{c|}{82.47}          & \multicolumn{1}{c|}{-}              & \multicolumn{1}{c|}{-}           & 85.61          \\
Q2ATransformer \cite{liu2023q2atransformer}              & \multicolumn{1}{c|}{}                                                               & \multicolumn{1}{c|}{-}            & \multicolumn{1}{c|}{81.20}          & \multicolumn{1}{c|}{-}              & \multicolumn{1}{c|}{-}            & 88.5           \\
MMBERT \cite{khare2021mmbert}                      & \multicolumn{1}{c|}{}                                                               & \multicolumn{1}{c|}{-}            & \multicolumn{1}{c|}{77.90}          & \multicolumn{1}{c|}{-}              & \multicolumn{1}{c|}{78.10} & -              \\
PubMedCLIP \cite{eslami2023pubmedclip}                  & \multicolumn{1}{c|}{}                                                               & \multicolumn{1}{c|}{-}            & \multicolumn{1}{c|}{80.00}          & \multicolumn{1}{c|}{82.50}          & \multicolumn{1}{c|}{-}           & -              \\
BiomedGPT-M \cite{zhang2023biomedgpt}                 & \multicolumn{1}{c|}{}                                                               & \multicolumn{1}{c|}{-}            & \multicolumn{1}{c|}{65.07}          & \multicolumn{1}{c|}{86.80}          & \multicolumn{1}{c|}{-}           & 85.70          \\
Prefix T. Medical LM \cite{van2023open}        & \multicolumn{1}{c|}{\textcolor{forestgreen}{\Checkmark}}                                  & \multicolumn{1}{c|}{1.5B}         & \multicolumn{1}{c|}{-}              & \multicolumn{1}{c|}{82.01}          & \multicolumn{1}{c|}{-}           & 87.00          \\
Gemini Pro \cite{qi2023gemini}                  & \multicolumn{1}{c|}{\textcolor{forestgreen}{\Checkmark}}                                  & \multicolumn{1}{c|}{-}            & \multicolumn{1}{c|}{60.29}          & \multicolumn{1}{c|}{72.60}          & \multicolumn{1}{c|}{60.22}       & 70.30          \\
\midrule
\multicolumn{7}{l}{Supervised finetuning results} \\
\midrule
MedThink \cite{gai2024medthink}                   & \multicolumn{1}{c|}{}                                                               & \multicolumn{1}{c|}{223M}         & \multicolumn{1}{c|}{83.50}          & \multicolumn{1}{c|}{86.30}          & \multicolumn{1}{c|}{-}           & 87.00          \\
LLaVA \cite{liu2024visual}                       & \multicolumn{1}{c|}{\textcolor{forestgreen}{\Checkmark}}                                  & \multicolumn{1}{c|}{7B}           & \multicolumn{1}{c|}{65.07}          & \multicolumn{1}{c|}{63.22}          & \multicolumn{1}{c|}{-}           & 63.20          \\
LLaVA-Med(LLama)  \cite{li2024llava}      & \multicolumn{1}{c|}{\textcolor{forestgreen}{\Checkmark}}                                  & \multicolumn{1}{c|}{7B}           & \multicolumn{1}{c|}{\underline{84.19}} & \multicolumn{1}{c|}{85.34}          & \multicolumn{1}{c|}{-}           & 91.21          \\
LLaVA-Med(Vicuna) \cite{li2024llava}     & \multicolumn{1}{c|}{\textcolor{forestgreen}{\Checkmark}}                                  & \multicolumn{1}{c|}{7B}           & \multicolumn{1}{c|}{81.98}          & \multicolumn{1}{c|}{83.17}          & \multicolumn{1}{c|}{-}           & \textbf{91.65} \\
LLaVA-Med(Phi2)  \cite{li2024llava}       & \multicolumn{1}{c|}{\textcolor{forestgreen}{\Checkmark}}                                  & \multicolumn{1}{c|}{2.7B}         & \multicolumn{1}{c|}{81.35}          & \multicolumn{1}{c|}{83.29}          & \multicolumn{1}{c|}{-}           & 90.17          \\
Med-MoE \cite{jiang2024med}                     & \multicolumn{1}{c|}{\textcolor{forestgreen}{\Checkmark}}                                  & \multicolumn{1}{c|}{2.0B}         & \multicolumn{1}{c|}{80.07}          & \multicolumn{1}{c|}{83.41}          & \multicolumn{1}{c|}{76.56}           & 91.30          \\
\rowcolor[HTML]{F3F3F3} V2T-CoT (Stablelm) & \multicolumn{1}{c|}{\textcolor{forestgreen}{\Checkmark}} & \multicolumn{1}{c|}{1.6B}         & \multicolumn{1}{c|}{80.08}          & \multicolumn{1}{c|}{\underline{86.48}} & \multicolumn{1}{c|}{\underline{79.69}} & 90.42          \\
\rowcolor[HTML]{F3F3F3} V2T-CoT(Phi2) & \multicolumn{1}{c|}{\textcolor{forestgreen}{\Checkmark}} & \multicolumn{1}{c|}{2.7B}         & \multicolumn{1}{c|}{\textbf{84.86}} & \multicolumn{1}{c|}{\textbf{87.61}} & \multicolumn{1}{c|}{\textbf{80.10}} & \underline{91.42} \\

\bottomrule
\end{tabular}}
\label{tab:medical_vqa_comparison}
\end{table*}

\subsection{Main Results}
In the experiment, we compared mainstream methods on four Med-VQA datasets, using accuracy as the metric for closed-end questions.
As shown in Table \ref{tab:medical_vqa_comparison}, 
V2T-CoT (Phi2) achieved the highest accuracy on VQA-RAD and SLAKE, demonstrating superior performance. 
A significant improvement over the baseline results can be observed, with an increase of 2.39\% on VQA-RAD and 5.11\% on SLAKE, respectively, outperforming LLaVA-Med (Vicuna) \cite{li2024llava}. 
Additionally, it attained competitive results on VQA-2019 and PathVQA, surpassing the compared method by 2.00\% over MMBERT \cite{khare2021mmbert} and 1.25\% over LLaVA-Med (Phi2) \cite{li2024llava}, respectively.
For open-end questions, Fig.~\ref{pipeline4}\textcolor{red}{A} provides a comparison of the performance of V2T-CoT and MedThink \cite{gai2024medthink} across three datasets, using open-ended evaluation metrics such as BLEU and Rouge-L. 
Results show that V2T-CoT consistently outperforms MedThink across all metrics in three datasets. 
Specifically, V2T-CoT highlights superior performance in BLEU-1 and Rough-L, highlighting its enhanced understanding and generation capabilities.

\begin{table}[htbp]
\large
\setlength{\abovecaptionskip}{10pt} 
\centering
\caption{Ablation Study on the Impact of Vision and Text CoT Integration: Performance Comparison on Closed-end and Open-end Questions. The \textbf{bolded} and \underline{underlined} values correspond to the best and 2nd best performing results.}
\resizebox{0.96\textwidth}{!}{ % Adjust table width to the page width
\begin{tabular}{cc|ccc|ccc}
\toprule
\multirow{2}{*}{Vision CoT} & \multirow{2}{*}{Text CoT} & \multicolumn{3}{c|}{\textbf{Closed-end Question (Accuracy)}} & \multicolumn{3}{c}{\textbf{Open-end Question (Recall)}} \\
                             &                           & {VQA-RAD} & {SLAKE} & {PathVQA} & {VQA-RAD} & {SLAKE} & {PathVQA} \\ 
\midrule
                             &                           & \underline{83.82}          & 84.38         & \underline{91.33}            & 56.45          & 82.73        & 29.75           \\
                             
                             & \textcolor{forestgreen}{\Checkmark} & 82.34          & 86.48        & 90.83          & 58.72          & \underline{83.24}        & \underline{31.57}          \\
                             \textcolor{forestgreen}{\Checkmark} &                           & 83.27          & \underline{87.04}         & \underline{91.33}            & \underline{59.53}          & 80.24        & 29.70           \\
\textcolor{forestgreen}{\Checkmark}  & \textcolor{forestgreen}{\Checkmark} & \textbf{84.86}          & \textbf{87.61}        & \textbf{91.42}           & \textbf{60.40}          & \textbf{84.08}        & \textbf{31.62}          \\
\bottomrule
\end{tabular}
}
\label{tab:ablation analysis}
\end{table}
\subsection{Ablation Analysis}

Table \ref{tab:ablation analysis} demonstrates the necessity of integrating visual and textual reasoning chains in V2T-CoT. 
V2T-CoT outperforms all datasets, showcasing the complementary roles of its components.

Vision CoT improves spatial reasoning, reducing SLAKE’s closed-end question error by 2.66\%.

Text CoT is crucial for logical coherence in open-ended tasks, with its absence causing a 2.27\% drop in VQA-RAD open-ended recall.

\subsubsection{Case Study Analysis.} 
\begin{figure}[htbp]
    \centering
    \includegraphics[width = \textwidth]{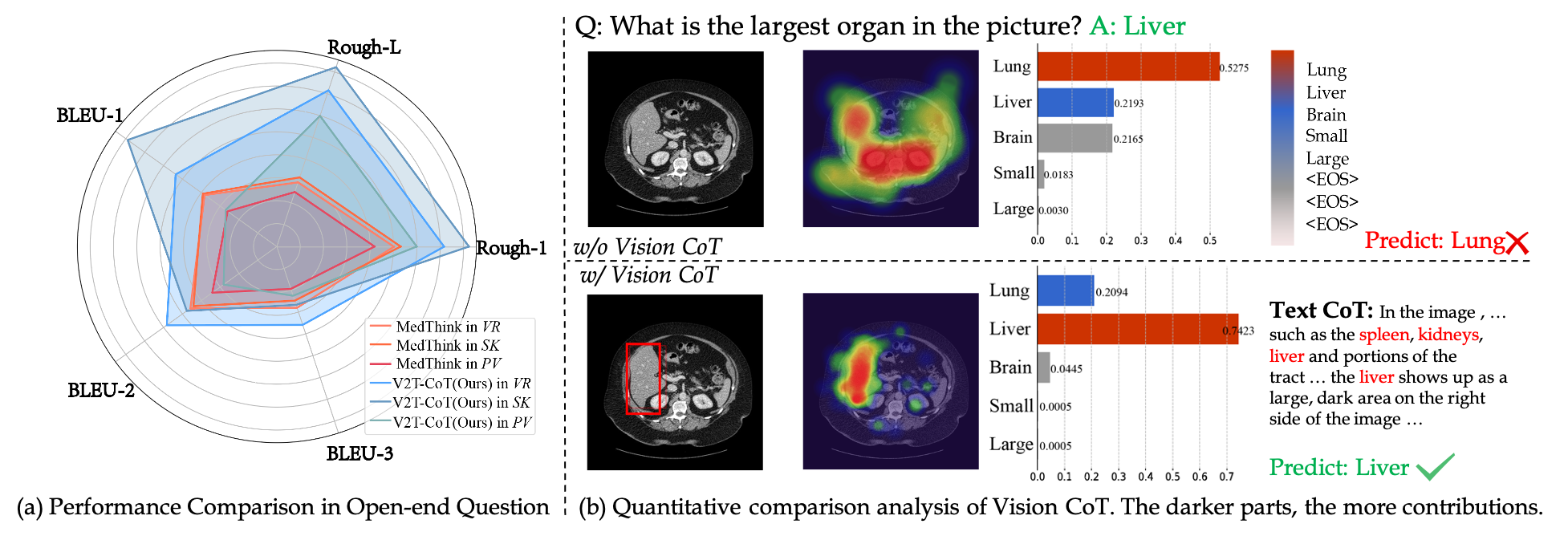}
    \caption{(a) A radar plot compares the performance of V2T-CoT and MedThink across three datasets: VQA-RAD(\textit{VR}), SLAKE(\textit{SK}), PathVQA(\textit{PV}) using open-ended evaluation metrics. (b) A quantitative heatmap visualization showing that incorporating Vision CoT improves attention to the organ (liver).}
    \label{pipeline4}
\end{figure}

Fig.~\ref{pipeline4}\textcolor{red}{B} demonstrates Vision CoT’s capability in lesion-centric attention guidance, where heatmap precisely localize anatomical structures (e.g., liver identification as the largest organ), thereby improving reasoning accuracy and diagnostic interpretability through targeted visual grounding.

\subsubsection{Vision CoT Regional Detection Analysis.}
To evaluate the impact of Vision CoT’s localization capability on Med-VQA task, we compare various object detection algorithms and Vision CoT variants: $\text{VCoT}_{G_T}$(Tiny)  and $\text{VCoT}_{G_L}$ (Large). Using mean Average Precision (mAP) as the evaluation metric for object detection, Table~\ref{tab:performance} demonstrates a positive correlation between VQA accuracy and mAP. 
Results indicate that Vision CoT variants consistently enhance diagnostic performance, with higher mAP directly contributing to improved medical diagnosis accuracy.
In Table~\ref{tab:anatomy}, we conducted cross-region validation on five anatomical regions from SLAKE to evaluate the impact of organ-specific annotations in $\text{VCoT}_{G_L}$ on VQA performance. Results show that mAP strongly correlates with VQA accuracy (Pearson’s r=0.92), highlighting Vision CoT’s ability to integrate anatomical-aware attention for improved diagnostic reasoning.

\begin{center}
% 第一个表格
\begin{minipage}[t]{0.58\textwidth}
\centering
\footnotesize  % 方法1：缩小基础字号
\setlength{\tabcolsep}{6pt} % 方法2：压缩列间距
\begin{tabularx}{\linewidth}{>{\raggedright\arraybackslash}c|ccc} % 方法3：优化对齐
\toprule
         & RTM-Det & $\text{VCoT}_{G_T}$       &  $\text{VCoT}_{G_L}$      \\
mAP(\%)  & \cellcolor{gray!15} 42.20   & \cellcolor{gray!25} 51.70        & \cellcolor{gray!35}61.00        \\ \midrule
\scriptsize VQA-RAD  & 77.57   & 80.88$_{\textcolor{red}{+3.31}}$ & 84.86$_{\textcolor{red}{+7.29}}$ \\
SLAKE    & 82.83   & 83.41$_{\textcolor{red}{+0.58}}$ & 87.61$_{\textcolor{red}{+4.78}}$ \\
PathVQA  & 90.50   & 90.83$_{\textcolor{red}{+0.33}}$ & 91.30$_{\textcolor{red}{+0.80}}$ \\
Med-2019 & 78.10   & 78.10$_{\textcolor{red}{+0.00}}$ & 80.10$_{\textcolor{red}{+2.00}}$ \\ \bottomrule
\end{tabularx}
\captionof{table}{Comparison of different Vision Detection method performance and the effect on Related Med-VQA.}
\label{tab:performance}
\end{minipage}%
\hfill
% 第二个表格
\begin{minipage}[t]{0.38\textwidth}
\centering
\footnotesize  % 统一字号设置
\setlength{\tabcolsep}{3.5pt}
\begin{tabularx}{\linewidth}{>{\raggedright\arraybackslash}c|>{\raggedleft}c>{\raggedleft\arraybackslash}c}
\toprule
Regions              & mAP(\%) & Acc(\%)  \\ \midrule
Abdomen       & 62.43   & 84.43       \\
Brain         & 69.05   & 88.00          \\
Lung          & 65.33   & 91.60       \\
Neck          & 42.23   & 68.75       \\
Pelvic Cavity & 51.74   & 68.42       \\ \bottomrule
\end{tabularx}
\captionof{table}{mAP and accuracy for different body regions on SLAKE dataset.}
\label{tab:anatomy}
\end{minipage}
\end{center}

\begin{figure}[!htbp]
    \centering
    \includegraphics[width = \textwidth]{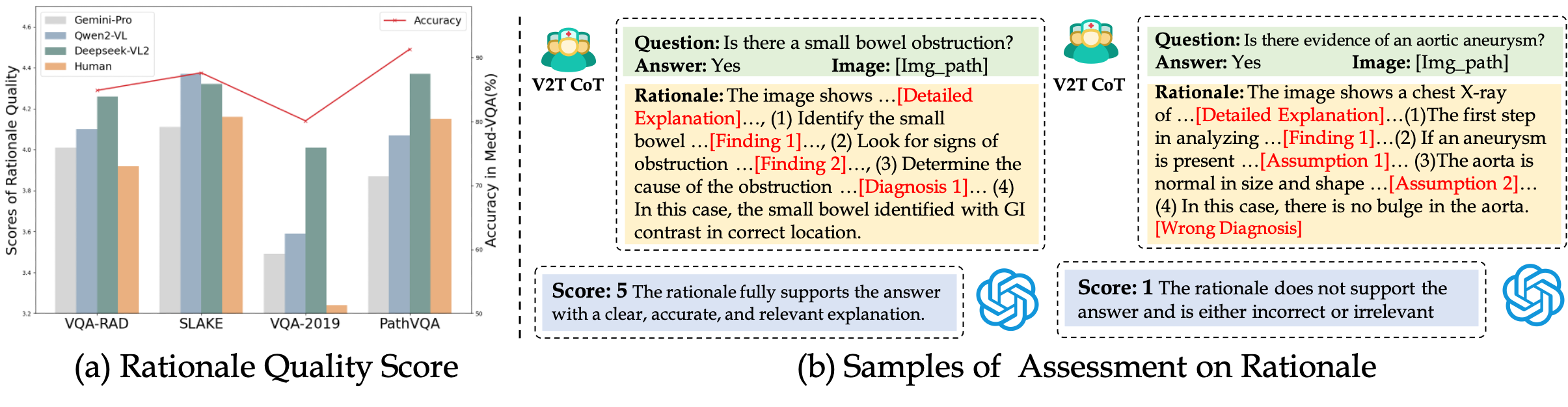}
    \caption{
    LLM and Human evaluation for rationale from \textbf{R-Med 39K}. 
    % for various models across medical VQA datasets. 
    (a) A bar chart illustrating rationale quality scores using different LLMs and human on four Med-VQA datasets, highlighting that higher rationale quality correlates with better Med-VQA performance.
    (b) Sample rationale quality assessments demonstrating Text CoT's role in medical diagnostics.}
    \label{fig:judgement}
\end{figure}
\noindent\textbf{Text CoT Rationale Analysis.}
\label{Text CoT Rationale}
In Fig.~\ref{fig:judgement}, we assess four Med-VQA datasets through LLM-based automated scoring (1–5 scale) and  human manual validation, focusing on rationale clarity in diagnostic reasoning. 
Results show VQA-RAD and SLAKE achieve average scores exceed 4, with comparable performance in VQA-2019 and PathVQA (Fig.~\ref{fig:judgement}\textcolor{red}{A}). The evaluation trend positively correlates with medical diagnosis accuracy, validating our assessment metric. Case studies (Fig.~\ref{fig:judgement}\textcolor{red}{B}) contrast a Score 5 rationale with explicit reasoning against a Score 1 case showing hallucinated assumptions and diagnostic errors. This demonstrates Text CoT’s ability to improve reliability via interpretable diagnostic reasoning.

\section{Conclusion}
In this paper, we propose V2T-CoT, a novel multimodal reasoning framework that significantly enhances the accuracy and interpretability of Med-VQA by Vision and Text CoT reasoning. 
Vision CoT locates disease-related visual cues, while Text CoT, trained on the constructed R-Med 39K instruction-tuning dataset, provides reasoning paths to enhance accuracy, interpretability, and transparency in the medical reasoning and diagnosis.

\begin{credits}
\subsubsection{\ackname} This work is supported by the National Natural Science Foundation of China (Grant No. 
62476241), the Natural Science Foundation of Zhejiang Province, China (Grant No. LZ23F020008),
and the Zhejiang University-Angelalign Inc. R\&D
Center for Intelligent Healthcare.

\end{credits}

\bibliographystyle{splncs04}

\end{document}